\documentstyle[12pt]{article} 
\input amssym.def 
\input amssym 
\newcommand{\be}{\begin{equation}} 
\newcommand{\ee}{\end{equation}} 
\begin{document} 

\title{Anti-de Sitter-type 2+1 spacetime of a charged 
rotating mass} 
\author{Nikolai V. Mitskievich\footnote{Departamento 
de F{\'\i}sica, CUCEI, Universidad de Guadalajara, 
Guadalajara, Jalisco, M\'exico; private postal address: 
Apartado Postal 1-2011, 44100 Guadalajara, Jalisco, 
M\'exico; e-mail: nmitskie@udgserv.cencar.udg.mx} 
\,and Alberto A. Garc{\'\i}a\footnote{Departamento 
de F{\'\i}sica, CINVESTAV del IPN, Apartado Postal 
14-740, 07000 M\'exico, D.F., M\'exico; e-mail: 
aagarcia@fis.cinvestav.mx}} 
\date{~} 
\maketitle 
\begin{abstract} 
The exact charged rotating solution of 2+1 Einstein--Maxwell
equations with $\Lambda$ term is obtained and its
properties outlined. It generalizes the
Cataldo--Cruz--del Campo--Garc{\'\i}a
relativistic charged massive black hole on the 2+1 anti-de
Sitter cosmological background. We show that rotating
solutions correspond to inhomogeneous field equations,
thus presence of sources in 2+1 Maxwell's equations cannot
be identified with existence of a charge distribution. Instead,
these sources are related to the 2+1 Machian 2-form field,
and the overall Lagrangian structure of the rotating system
is reconstructed.
\end{abstract} 

In 2+1 spacetime the 1-form field has to obey nonlinear 
equations to yield zero-trace stress-energy tensor (we 
call this the intrinsically relativistic field condition). 
This automatically guarantees a profound analogy of its 
spherically symmetric static electrovacuum solution with 
the 3+1 Reissner--Nordstr\"om--Kottler solution, in particular 
existence of asymptotic infinities and horizons. Such a 
solution was first discovered and studied in \cite{CatGar} 
(C$^3$G) [below we refer to formulae in this paper as, for 
example, to (1C$^3$G)]. 

Here we find that there exists a simple generalization of the 
static C$^3$G solution to the case of a rotating central mass 
possessing similar geometric and physical properties, as 
well as being analogous to the Kerr--Newman spacetime 
with a cosmological term. Since the new solution is a close 
analogue of the C$^3$G one and its geometric properties are 
fairly similar to those of the latter, we shall not consider in 
this paper the horizons and infinities in this spacetime, 
but concentrate instead on field theoretic problems.

We shall use a 2+1-dimensional spacetime with signature 
$+$ $-$ $-$ (thus det\,$g_{\mu\nu}>0$) and the definition 
of the Ricci tensor as $R_{\mu\nu}={R^\alpha}_{\mu\nu\alpha}$. 
Then Einstein's equations take the form 
\be 
R_{\mu\nu}-\frac{1}{2}g_{\mu\nu}R=-\varkappa T_{\mu\nu}+ 
\Lambda g_{\mu\nu},    \label{Einst} 
\ee 
while $R=-6\Lambda$ when the trace $T_{\mu\nu}g^{\mu\nu}$ 
vanishes, hence in our case 
\be 
R_{\mu\nu}=-\varkappa T_{\mu\nu}-2\Lambda g_{\mu\nu}. 
\ee 
For a 1-form field $A$ with the skew-symmetric field 
tensor $F=dA$ this means that the Lagrangian density for 
this field is ${\frak L}=\sqrt{g}L$, $L=\sigma(-I)^{3/4}$ where 
$I=F_{\mu\nu}F^{\mu\nu}/2$ is the only first-order differential 
1-form field invariant in 2+1. In the case under consideration 
(the field produced by a rotating charge), $I<0$. 

Taking now the Kerr metric in the Boyer--Lindquist coordinates 
and fixing the polar angle $\vartheta=\pi/2$, we obtain 
\be 
ds^2=\frac{\Delta-a^2}{r^2}\left(dt+ 
a\frac{r^2+a^2-\Delta}{\Delta-a^2}d\phi\right)^2- 
\frac{r^2}{\Delta}dr^2-\frac{r^2\Delta}{\Delta-a^2}d\phi^2, 
     \label{ds2} 
\ee 
$\sqrt{g}=r$. When $a=0$, this squared interval takes the 
C$^3$G form for $\Delta_{{\rm C^3G}}=-Mr^2+\Lambda 
r^4+Q^2r$, so that we only have to add to it $a^2$ in order 
to arrive at a (still hypothetical) rotating generalization: 
\be 
\Delta(r)=a^2+Q^2r-Mr^2+\Lambda r^4.    \label{Delta} 
\ee 
It is important to take into account the fact that the 
dimensionalities of the constants $Q$ and $M$ are here 
(as well as in \cite{CatGar}) different from those 
in the Reissner--Nordstr\"om or Kerr--Newman metric: 
the (topological) mass $M$ is dimensionless, and the 
(nonlinear) charge $Q^2$ has the dimension of length. 

In the natural orthonormal triad $\theta^{(\mu)}$ 
\be   \label{triad} 
\theta^{(0)}=\frac{\sqrt{\Delta-a^2}}{r}\left(dt+a\frac{r^2+ 
a^2-\Delta}{\Delta-a^2}d\phi\right), ~ \theta^{(1)}= 
\frac{r}{\sqrt{\Delta}}dr, ~ \theta^{(2)}=\frac{r\sqrt{\Delta} 
}{\sqrt{\Delta-a^2}}, 
\ee 
corresponding to the quadratic form (\ref{ds2}), the 
Ricci tensor has non-zero independent components 
$$ 
R_{(0)(0)} =-\frac{Q^2}{2r^3}\left(1+\frac{3a^2}{\Delta- 
a^2}\right)-2\Lambda, 
$$ 
$$ 
R_{(1)(1)} =\frac{Q^2}{2r^3}+2\Lambda, 
$$ 
$$ 
R_{(2)(2)} =-\frac{Q^2}{2r^3}\left(2+\frac{3a^2}{\Delta- 
a^2}\right)+2\Lambda, 
$$ 
$$ 
R_{(0)(2)} =\frac{3aQ^2\sqrt{\Delta}}{2r^3\left(\Delta 
-a^2\right)} 
$$ 
(so that the Ricci scalar is $R=-6\Lambda$), 
thus we find from Einstein's equations (\ref{Einst}) 
$$ 
\varkappa T_{(0)(0)} =\frac{Q^2}{2r^3}\left(1+ 
\frac{3a^2}{\Delta-a^2}\right), 
$$ 
$$ 
\varkappa T_{(1)(1)} =-\frac{Q^2}{2r^3}, 
$$ 
$$ 
\varkappa T_{(2)(2)} =\frac{Q^2}{2r^3}\left(2+ 
\frac{3a^2}{\Delta-a^2}\right), 
$$ 
$$ 
\varkappa T_{(0)(2)} =-\frac{3aQ^2\sqrt{\Delta}}{2r^3 
\left(\Delta-a^2\right)} 
$$ 
(for triad components, indices in parentheses are used, 
the triad metric being 2+1 Minkowskian with the signature 
$+$ $-$ $-$, $g_{(\mu)(\nu)}=g^{(\mu)(\nu)}$). We see 
that the stress-energy tensor is traceless indeed. 

Eigenvalues of the stress-energy tensor (here we take  
this tensor with one co- and one contravariant indices) are: 
$\lambda_1=\lambda_2=Q^2/(2\varkappa r^3)$ and 
$\lambda_3=-Q^2/(\varkappa r^3)$. It is characteristic that 
these eigenvalues do not contain any information about 
the mass $M$ and angular momentum $a$. The 
corresponding eigenvectors $e^{(\mu)}$ (let us consider 
them as covectors with the components 
${e^{(\mu)}}_{(\alpha)}$ with respect to the usual basis 
$\theta^{(\alpha)}$, thus $e^{(\mu)}={e^{(\mu)}}_{(\alpha)} 
\theta^{(\alpha)}$) take in terms of $\theta^{(\mu)}$ the form 
\be 
e^{(0)}=N\left(\sqrt{\Delta}\theta^{(0)}-a\theta^{(2)}\right), ~ 
e^{(1)}=\theta^{(1)}, ~ 
e^{(2)}=N\left(-a\theta^{(0)}+\sqrt{\Delta}\theta^{(2)}\right) 
\ee 
where the normalization factor is $N=\left(\Delta-a^2\right)^{-1/2}$. 
The triad $e^{(\mu)}$ is, naturally, orthonormal due to the 
symmetry of the stress-energy tensor. The inverse 
transformation reads 
\be 
\theta^{(0)}=N\left(\sqrt{\Delta}e^{(0)}+ae^{(2)}\right), ~ 
\theta^{(1)}=e^{(1)}, ~ 
\theta^{(2)}=N\left(ae^{(0)}+\sqrt{\Delta}e^{(2)}\right) 
\ee 
with the same $N$. A substitution of the expressions (\ref{triad}),
after a simple coordinate change $t\rightarrow t+a\phi$, yields
the orthonormal covector basis
\be   \label{etriad} 
e^{(0)}=\frac{\sqrt{\Delta}}{r}dt, ~ e^{(1)}=\frac{r}{\sqrt{\Delta}}dr, 
~ e^{(2)}=r\left(d\phi-\frac{a}{r^2}dt\right), 
\ee 
so that the metric tensor becomes much simpler than it 
was previously, (\ref{ds2}): 
\be    \label{emetric} 
ds^2=\frac{\Delta}{r^2}dt^2-\frac{r^2}{\Delta}dr^2- 
r^2\left(d\phi-\frac{a}{r^2}dt\right)^2. 
\ee
We also see that the vector basis corresponding to
(\ref{etriad}) is
\be \label{vetriad}
X_{(0)}=\frac{r}{\sqrt{\Delta}}\left(\partial_t+\frac{a}{r^2}
\partial_\phi\right), ~ X_{(1)}=\frac{\sqrt{\Delta}}{r}\partial_r,
~ X_{(2)}=\frac{1}{r}\partial_\phi.
\ee

In the new basis, the stress-energy tensor components
take the following form:
$$
T_{(0)(0)} =\frac{Q^2}{2\varkappa r^3}=\lambda_1,
$$
$$
T_{(1)(1)} =-\frac{Q^2}{2\varkappa r^3}=-\lambda_1,
$$
$$
T_{(2)(2)} =\frac{Q^2}{\varkappa r^3}=-\lambda_3
$$
(the non-diagonal components are equal to zero).
This makes it clear that the respective orthonormalized
eigenvectors of $T^{(\alpha)}_{(\beta)}$ are exactly
$e^{(0)}$, $e^{(1)}$ and $e^{(2)}$. Now the vanishing
of the trace $T=T_{(\mu)(\nu)}g^{(\mu)(\nu)}$ is quite
obvious.

While the fulfilment of Einstein's equations guarantees
fulfilment of the field equations of the 1-form field
(usually labelled as ``electromagnetic'') since the exact
gravitational field theory is automatically self-consistent,
it would still be worth considering these 1-form field
equations to the end of better understanding of this
field's Lagrangian structure. The first (and na{\"\i}ve)
idea coming to the mind is to simply use the Lagrangian
density ${\frak L}=\sqrt{g}L(I)$ where $I=\frac{1}{2}
F_{\mu\nu}F^{\mu\nu}$, $F=\frac{1}{2}F_{\mu\nu}
dx^\mu\wedge dx^\nu=dA$, $A=A_\mu dx^\mu$ being
the 1-form field potential. But in a 2+1-dimensional
spacetime there exists an opportunity to introduce the
1-form $f$ equivalent to 2-form $F$ as
\be
\left\{ ~f_\lambda=\frac{1}{2}E_{\lambda\mu\nu}F^{\mu\nu},
~ F^{\mu\nu}=E^{\lambda\mu\nu}f_\lambda ~ \right\} ~
\Leftrightarrow ~ \left\{ ~ f=\ast F, ~ F=\ast f ~ \right\},
\ee
the Levi-Civit\`a axial tensor being
\be
E_{\lambda\mu\nu}=\sqrt{g}\epsilon_{\lambda\mu\nu},
~ ~ E^{\lambda\mu\nu}=\frac{1}{\sqrt{g}}\epsilon_{\lambda
\mu\nu},
\ee
while
\be
\epsilon_{\lambda\mu\nu}=\epsilon_{[\lambda\mu\nu]}, ~ ~
\epsilon_{012}=+1
\ee
is the Levi-Civit\`a constant three-dimensional symbol (the
antisymmetrization in a group of indices is denoted, as
usual, with the Bach brackets on the both sides of them);
$\ast$ is the 2+1 Hodge star ($\ast\ast=+1$).

In order to incorporate into our description the rotation
property of the new solution, let us write the field
equations as
\be     \label{1formeq}
\left(\sqrt{g}\frac{dL}{dI}F^{\mu\nu}\right)_{,\nu}=-4\pi
\sqrt{g}j^\mu
\ee
which are equivalent to
\be   \label{1aformeq}
d\left(\frac{dL}{dI}f\right)=-4\pi\ast j.
\ee
It is known from \cite{CatGar} that the 2+1 nonrotating
intrinsically relativistic 1-form field is described by the
Lagrangian $L=\sigma(-I)^{3/4}$, (6C$^3$G), where
$\sigma=$ const. In this case, the ``source term''
containing $j$ will be absent, and
the equations (\ref{1formeq}) and (\ref{1aformeq})
become ``homogeneous'' (in fact, as this will be seen
below, the ``inhomogeneity'' --- when it is present ---
consists of two terms one of which is proportional to
$A$ and another, to $F$, so that its characterization
as an inhomogeneity is not completely adequate).  It is
clear that $j=0$ manifests the absence of rotation of
the 1-form $f$ congruence. Thus, when $a=0$, the
squared interval (\ref{emetric}) reduces to that found
in \cite{CatGar}, and the equations (\ref{1formeq}), to
(4C$^3$G), without the ``source'' term. However, as we
shall see, the presence of rotation changes the
situation drastically. In particular, this change cannot
be cast into the Lagrangian form without involvement
of a field whose potential is a 2-form (in 2+1).

From the general expression
\be
T^\beta_\alpha=\left(2I\frac{dL}{dI}-L\right)\delta^\beta_\alpha
-2\frac{dL}{dI}f_\alpha f^\beta
\ee
it can be seen that $f$ is the eigenvector corresponding
to the single eigenvalue $\lambda_3$. Thus $f$ has to be
proportional to $e^{(2)}$ which rotates, as it follows from
(\ref{etriad}),
\be
f=\frac{Q^{4/3}}{(\varkappa\sigma)^{2/3}r^2}e^{(2)},
\ee
$f\wedge df\neq 0$, while
\be   \label{F}
F=dA=\frac{Q^{4/3}}{(\varkappa\sigma)^{2/3}r^2}e^{(0)}
\wedge e^{(1)}=\frac{Q^{4/3}}{(\varkappa\sigma)^{2/3}r^2}
dt\wedge dr,
\ee
\be   \label{A}
A=\frac{Q^{4/3}}{(\varkappa\sigma)^{2/3}
\sqrt{\Delta}}e^{(0)}=\frac{Q^{4/3}}{(\varkappa\sigma)^{2/3}r}dt,
\ee
\be
I=f_\mu f^\mu=-\frac{Q^{8/3}}{(\varkappa\sigma)^{4/3}r^4}, ~
~ \frac{dL}{dI}=-\frac{3\sigma(\varkappa\sigma)^{1/3}r}{4
Q^{2/3}}.
\ee
Thus the expression being differentiated in (\ref{1aformeq})
does not represent an exact or closed 1-form due to the
rotation parameter $a$ different from zero, as this was to
be expected:
\be   \label{rotation}
\frac{dL}{dI}f=-\frac{3(\sigma Q)^{2/3}}{4\varkappa^{1/3}}
\left(d\phi-\frac{a}{r^2}dt\right)=-\frac{3(\sigma Q)^{2/3}}{4
\varkappa^{1/3}r}e^{(2)}.
\ee

It is now obvious that when a rotating field is to be
considered, the ``homogeneous'' variants of field
equations (\ref{1formeq}) and ({\ref{1aformeq}) have
to be revised (this fact makes it clear why the approach
to rotating black holes in \cite{Banetal} is wrong, thus
supporting the criticisms given in \cite{Garcia}). This
can be done if an additional term would be incorporated
into the Lagrangian. The situation is essentially the
same as that discovered for a general 3+1 field
theoretic description of perfect fluids in \cite{Mits99a,
Mits99b}. The new term has to leave the stress-energy
tensor unchanged, thus ({\em cf.} \cite{Mits99a} with a
necessary modification for the present dimensionality of
spacetime) we choose this term to be proportional to the
(invariant) combination of completely skew rank-three
tensors. Their dependence on the already used objects
does not matter (in the spirit of the Noether theorem).

Thus we choose two such tensors: the antisymmetrized
product $A_{[\lambda}F_{\mu\nu]}$ [in fact, vanishing
due to (\ref{F}) and (\ref{A}), but giving nontrivial
contributions when differentiated with respect to $A$
and $F$], and the 2-form field's field tensor (3-form),
\be
W=dC, ~ ~ W_{\lambda\mu\nu}=\tilde{W}E_{\lambda
\mu\nu}
\ee
(the Machian or cosmological field, {\em cf.} for 3+1
\cite{Mits99a}). These two objects form the invariant
\be
J=W^{\lambda\mu\nu}A_\lambda F_{\mu\nu}\equiv
2\tilde{W}A_\lambda f^\lambda
\ee
(being equal to zero for the solution under consideration),
which we use to construct the additional term $M$ in the
Lagrangian: ${\frak L}=\sqrt{g}(L(I)+M(J))$.
The stress-energy tensor corresponding to $M$ is
\be    \label{ghostT}
T^\beta_\alpha=\left(2J\frac{dM}{dJ}-M\right)
\delta^\beta_\alpha;
\ee
the new 1-form field equation takes the form
\be   \label{rot1f}
d\left(\frac{dL}{dI}f+\frac{dM}{dJ}\tilde{W}A\right)=
-\frac{dM}{dJ}\tilde{W}F,
\ee
and the (formally) 2-form field equation,
\be
d\left(\frac{dM}{dJ}A_\lambda f^\lambda\right)=0.
\ee
The only case when the last equation is fulfilled
identically while the equation (\ref{rot1f}) provides
the possibility of rotation, is realised when $M=J$.
Then $M$ vanishes due to the properties of our
solution, so that the 2-form field's stress-energy
tensor (\ref{ghostT}) vanishes too (the ghost
property with respect to Einstein's equations,
but not to the Maxwell-type ones); from (\ref{rot1f})
it follows that
\be   \label{Wrot}
d\left(\frac{dL}{dI}f\right)+d\tilde{W}\wedge A
+2\tilde{W}F=0,
\ee
yielding
\be
\tilde{W}=Sr^2+\frac{aH}{2r},
 ~ ~ S={\mbox{ (arbitrary) const.}}, ~ ~ H=\sigma
\left(\frac{\varkappa\sigma}{Q^2}\right)^{1/3}.
\ee
This completes the explicit proof of self-consistency
of our solution, although the fact of fulfilment of
Einstein's equations already could not leave any
room for doubts.

Since it is more common to use the field equation 
in the form (\ref{1formeq}), with a non-zero right-hand
side in the rotating case, we now calculate
$$
\sqrt{g}\frac{dL}{dI}F^{\mu\nu}=\frac{3(\sigma Q)^{2/3}
}{4\varkappa^{1/3}}\left(\epsilon^{\phi\mu\nu}-
\frac{a}{r^2}\epsilon^{t\mu\nu}\right)
$$
$$
\equiv-\frac{3a(\sigma Q)^{2/3}}{4\varkappa^{1/3}r^2}
\left(\delta^\mu_r\delta^\nu_\phi-\delta^\mu_\phi
\delta^\nu_r\right)+{\mbox{ const}}^{\mu\nu}.
$$
Thus
\be   \label{j}
j=\frac{3a(\sigma Q)^{2/3}}{8\pi\varkappa^{1/3}r^3}
X_{(2)}
=\frac{3a(\sigma Q)^{2/3}}{8\pi\varkappa^{1/3}r^4}
\partial_\phi.
\ee
We see that the source term, usually interpreted as
the electric current density, is spacelike (hence not
attributable to any moving charges). It is proportional
to the rotation parameter $a$, though it also contains
the coefficient $Q^{2/3}$, but the latter is related to
the central pointlike charge and by no means to any
spatial distribution of charges and/or currents. We
interpret the role of $Q$ as that of a ``catalyst'' of
rotation which otherwise (in the absence of an
``electric'' field) cannot manifest itself in the (then
absent) Maxwell-type equations. Moreover, the
``source'' term in (\ref{Wrot}) [or, equivalently, in
(\ref{1aformeq})] does linearly depend on the
1-form field potential and, in its other summand,
on the corresponding field tensor, --- the fact
being completely foreign to the basic ideas of the
conventional electromagnetic field theory (and there
is no other --- alternative --- solution to this problem).

The exact solution of a field generated by an obviously
pointlike source in 2+1-dimensional spacetime, when a
rotation is present, belongs to a distributed-source
containing system of Maxwell-type equations. This fact
leads to a radical revision of the physical interpretation
of distributed sources of 1-form fields in 2+1 ({\em not
as electric current densities}), as well as to a revision of
the physical meaning of these very fields (there is
essentially less analogy between a 1-form field in 2+1
and the electromagnetic one in 3+1 than it is usually
admitted. In particular, the ``magnetic'' type fields in
fact describe here 2+1 perfect fluids: {\em cf.} a
statement made in \cite{Mits99a, Mits99b} which should
be however somewhat modified in the ``electric'' case
just having been considered). The system
of fields describing a rotating massive charged centre
thus consists of the gravitational field ($g$), the 1-form
field ($A$), and the 2-form field [$C$, being a ghost
field in the sense of Einstein's equations, but inevitable
to fulfil the Maxwell-type equations (\ref{rot1f}) to
incorporate the rotation]. A more general and complete
treatment of the 1-form field theory in 2+1 will be found
in a subsequent publication.

\subsection*{Acknowledgements}
A.A.G. acknowledges a partial support from the CONACyT
(Mexico) Project 32138E; N.V.M. thanks CINVESTAV IPN
for hospitality during his visit there when this work was done.

\end{document}